\documentstyle[11pt]{article}
\begin{document}
\begin{center}
\renewcommand{\thefootnote}{\fnsymbol{footnote}}

\Large{\bf THE LARGEST VIRIALIZED DARK HALO IN THE UNIVERSE}\\[5mm]
WEI ZHOU, TONG-JIE ZHANG\footnote{E-mail: tjzhang@bnu.edu.cn}
, LI CHEN\\
 XIANG-TAO HE and YU-MEI HUANG\ \ \\
{\small \it Department of Astronomy, Beijing Normal University, Beijing, 100875,
China}\\[8mm]
\end{center}
\begin{flushleft}
Using semi-analytic approach, we present an estimate of the properties of
the largest virialized dark halos in the present universe for three
different scenarios of structure formation: SCDM, $\Lambda$CDM and
OCDM models.
The resulting virial mass and temperature increase from the lowest
values of
$1.6 \times 10^{15}h^{-1}M_{\odot}$ and 9.8 keV in OCDM, the
mid-range values of
$9.0 \times 10^{15}h^{-1}M_{\odot}$ and 31 keV in $\Lambda$CDM,
to the highest values of
$20.9 \times 10^{15}h^{-1}M_{\odot}$, 65 keV in SCDM.
As compared with the largest virialized object seen in the universe,
the richest clusters of galaxies, we can safely rule out the OCDM model.
In addition, the SCDM model is very unlikely because of the unreasonably
high virial mass and temperature. Our computation favors the
prevailing $\Lambda$CDM model in which superclusters
may be marginally regarded as the dynamically-virialized systems.
\\[8mm]
\end{flushleft}
\section{ Introduction}
\hspace{5mm} The central problem in modern cosmology is the
formation of large scale structures in the universe. In the
standard picture of hierarchical structure formation, dark matter
dominates the universe, and a wide variety of observed structures,
such as galaxies, groups and clusters of galaxies, have formed by
the gravitational growth of Gaussian primordial density
fluctuations. Due to self-gravitational instability, the
fluctuations of dark matter have collapsed and virialized into
objects which are so-called 'dark matter halos' or 'dark halos'.
The larger halos are generally considered to have formed via the
merger of smaller ones collapsed first. The distribution of mass
in the gravitationally collapsed structures, such as galaxies and
groups (or clusters) of galaxies, which is usually called the mass
or multiplicity function, has been determined by observation
.$^{1,2,3,4}$ The best approximate description for cosmological
mass function was proposed by Press and Schechter$^{5}$. The PS
theory did not draw much attention until 1988, when the first
relative large N-Body simulation$^{6,7}$ revealed a good agreement
with it. The mystery of the
 'fudge factor' of
2 in PS theory was solved by Peacock \& Heavens$^{8}$ and Bond
et al$^{9}$
 approaching the 'cloud-in-cloud' problem by a rigorous way. The
 reliability of the PS formula
has been tested using N-Body simulation by several authors, which
turns out the
PS formula indeed provides an overall satisfactory description of
mass function
for virialized objects. Among the virialized structures, galaxy
clusters are
extremely useful to cosmology because they may be in detail studied
as individual
objects, and especially are the largest virialized structure in the
universe at
present. The mass of a typical rich clusters is approximately
$10^{15}h^{-1}M_
{\odot}$, which is quite similar to the average mass within a
sphere of $8h^{-1}$Mpc
radius in the unperturbed universe. However, the theoretical
estimate of the
mass of the largest collapsed object in the cosmological framework
has still not
been presented. In this paper, we will calculate the mass function
of collapsed
objects by PS formula although it has already been carried out by
many authors.  Different
from the previous works, we will further focus on the derivation of
the mass of the largest virialized object by a volume integration
over the
whole Universe. In addition, the virial temperature and radius of
the largest structure are also
derived. \par
\section{ The Mass distribution of Virialized Objects}
\hspace{8mm} It is now believed that the comoving number density of
massive halos per unit mass
at a given redshift
$z$ and mass $M$ follows the PS formula
\begin{equation}
\frac{dn}{dM}=-\sqrt[]{\frac{2}{\pi}}\frac{\rho_{m}}{M}\frac
{d\ln\sigma(M)}{dM}\nu_{c}{\rm e}^
{-\frac{{\nu_{c}}^{2}}{2}}.
\end{equation}
where $\rho_{m}=\Omega_{m}\rho_{crit,0}$, is the present average
comoving matter density of the universe , the critical density at
the present $\rho_{crit,0}=3H_{0}^{2}/8\pi G=2.7755 \times
{10}^{11}h^{2}M_{\odot}$Mpc$^{-3}$ where the Hubble constant
$H_{0}=100h $km s$^{-1}$Mpc$^{-1}$ and the threshold function for
collapse $\nu_{c}=\delta_{crit}(z)/\sigma(M)$. In addition, we
introduce the cosmological density parameters \\
\begin{center}
$\Omega_{m}=\rho_{m}/\rho_{crit,0};\hspace{5mm}\Omega_{k}
=k/(H_{0}R_{0})^{2};\hspace{5mm}\Omega_{\Lambda}=\Lambda/(3H_{0}^{2})$\\
\end{center}
which satisfies $\Omega_{m}+\Omega_{\Lambda}-\Omega_{k}=1$ and
where $\Lambda$ is the cosmological constant, $R_{0}$ is today's
scale factor of the universe and $k=0,\pm 1$ is the spatial
curvature of the universe. The present mass variance for the
fluctuation spectrum fitted on mass $M$ takes the form
\begin{equation}
\sigma^{2}(M)=\frac{1}{2\pi^{2}}\int_{0}^{\infty}k^2p(k)W^{2}(Rk)\,{\rm d}k
\end{equation}
in which $W(x)=3(sinx-xcosx)/x^{3}$ works as the Fourier-Space
representation of a spherical top-hat window function and $M=4\pi
R^{3}\rho_{m}/3$. We parameterize the power spectrum of initial
fluctuation by
\begin{equation}
p(k)=Ak^{n}T^{2}(k)
\end{equation}
The normalization of the power spectrum is realized by $\sigma_{8}$, the rms
mass fluctuation on a scale of $8h^{-1}$Mpc, and the
Harrison-Zeldovich case of the primordial power spectrum, i.e., n=1,
is adopted. The transfer
function of an adiabatic CDM model is given by Bardeen et al.$^{10}$
\begin{equation}
T(q)=\frac{\ln(1+2.34q)}{2.34q}{[1+3.89q+{(16.1q)}^{2}+{(5.46q)}^{3}+{(6.71q)}
^{4}]}^{-\frac{1}{4}}
\end{equation}
where $q=(k/h$Mpc$^{-1})/\Gamma$, and $\Gamma=\Omega_{m}hexp[-\Omega_{b}
(1+\sqrt[]{2h}/\Omega_{m})]$ is the shape parameter with the baryon density
 $\Omega_{b}$, the mean density of baryons in the universe with respect to
 the critical density
 at the present.\par
The overdensity linearly extrapolated to the present epoch is characterized by
the normalized linear growth factor $D(z): \delta_{crit}(z)=\delta_{crit}
(0)/D(z)$,
where $\delta_{crit}(0)$ has a week dependence on $\Omega_{m}$ and can
 be approximated by
\begin{eqnarray}
\delta_{crit}(0)= \left \{
\begin{array}{ll}
0.15{(12\pi)}^{2/3} & \Omega_{k}=0 {\hspace{3mm}}  and {\hspace{3mm}}
\Omega_{\Lambda}=0  \\
0.15{(12\pi)}^{2/3}{\Omega_{m}}^{0.0185} & \Omega_{m}<1 {\hspace{3mm}}
and {\hspace{3mm}} \Omega_{\Lambda}=0 \\
0.15(12\pi)^{2/3}{\Omega_{m}}^{0.0055} & \Omega_{k}=0 {\hspace{3mm}}\\
\end{array}
\right.
\end{eqnarray}
where we have not taken into account the dependence of the
$\delta_{crit}(0)$ on redshift $z$, which Lokas and Hoffman even
consider$^{11}$. It is convenient to express the growth factor as
$D(z)=[g(z)/g(0)]/(1+z)$, where the best-fit of $g(z)$
reads$^{12}$
\begin{equation}
g(z)=\frac{(5/2)\Omega_{m}(z)}{{\Omega_{m}(z)}^{4/7}-\Omega_{\Lambda}(z)
+(1+\Omega_{m}(z)/2)(1+\Omega_{\Lambda}(z)/70)}
\end{equation}
where
\begin{equation}
\Omega_{m}(z)=\frac{\Omega_{m}(1+z)^{3}}{E^{2}(z)},{\hspace{3mm}}
\Omega_{\Lambda}(z)=\frac{\Omega_{\Lambda}}{E^{2}(z)}
\end{equation}
where $E^{2}(z)=\Omega_{m}(1+z)^{3}+\Omega_{k}(1+z)^{2}+\Omega_{\Lambda}$.
\\

\hspace{5mm}Based on the theoretical expression above, we can
easily get the total number $N$ of the virialized objects with the
mass larger than $M_{0}$
\begin{equation}
N=\int_{0}^{\infty}[\int_{M_{0}}^{\infty}\frac{dn}{dM}{\rm
d}M]\frac{dV}{dz}{\rm d}z \label{eq:nv}
\end{equation} where $dV$ is the comoving volume
element for the Friedmann-Robertson-Walker metric and
$\frac{dV}{dz}$ takes the
form\\
\begin{eqnarray}
\frac{dV}{dz}= \left \{
\begin{array}{ll}
4\pi\frac{c^{3}}{H_{0}^{3}}\frac{D_{a}^{2}(1+z)^{2}\cos[\sqrt{\Omega_{k}}
\cdot
f]}{\sqrt{1+\Omega_{k}D_{a}^{2}(1+z)^{2}}E(z)}
& for{\hspace{3mm}}\Omega_{k}<0 \\
4\pi\frac{c^{3}}{H_{0}^{3}}\frac{D_{a}^{2}(1+z)^{2}}{E(z)}& for{\hspace
{3mm}}\Omega_{k}=0 \\
4\pi\frac{c^{3}}{H_{0}^{3}}\frac{D_{a}^{2}(1+z)^{2}\cosh[\sqrt{\Omega_{k}}
\cdot
f]}{\sqrt{1+\Omega_{k}D_{a}^{2}(1+z)^{2}}E(z)}& for{\hspace{3mm}}
\Omega_{k}>0\\
\end{array}
\right.
\end{eqnarray}
where $D_{a}=d_{A}H_{0}/c$, $d_{A}$ is the angular diameter
distance and $f=\int_{0}^{z}{\rm d}z/E(z)$.
\section{ Results }
\hspace{8mm}In this paper, we exploit the three representative
cosmological
models: Model 1, standard CDM (SCDM) model, a flat model with
$\Omega_{m}=1$,
$\Omega_{\Lambda}=0$; Model 2, flat Lambda CDM
 ($\Lambda$CDM) model, a low density flat model with $\Omega_{m}=0.3$,
 $\Omega_{\Lambda}=0.7$;  Model 3, open CDM (OCDM) model , a low density
 open model with $\Omega_{m}=0.3$, $\Omega_{\Lambda}=0$, which are all
 in detail listed in Table 1. Throughout this paper, we take
 $H_{0}=100h$kms$^{-1}$Mpc$^{-1}$.\par
It is obvious from the Eq.(\ref{eq:nv}) that the total number $N$
decrease with the increase of the mass $M_{0}$. Setting $N=1$, we
can finally obtain the largest virial mass $M_{0}$ (or $M_{vir}$)
for virialized object, which is connected to the virial radius and
virial temperature respectively$^{13}$
\begin{equation}
M_{vir}=4 \pi r^{3}_{vir}\rho_{crit}\Delta_{c}/3,
\end{equation}
\begin{equation}
kT=1.39f_{T}(\frac{M}{10^{15}M_{\odot}})^{2/3}(h^{2}\Delta_{c}E^{2})
^{1/3}keV,
\label{eq:kt}
\end{equation}
where $\Delta_{c}$ represents the overdensity of dark matter with
respect to the critical density $\rho_{crit}(z)$ and can be
approximated by$^{14}$\\
\begin{center}
$\Delta_{c}=18\pi^{2}+82x-39x^{2}\hspace{3mm}$ for $\Omega_{k}=0$;
\hspace{5mm} $\Delta_{c}=18\pi^{2}+60x-32x^{2}\hspace{3mm}$ for
$\Omega_{\Lambda}=0$,
\end{center}
where $x=\Omega_{m}(1+z)^{3}/E^{2}(z)-1$.
The critical density of the universe at a given redshift $z$ is
related to the present critical average mass density by
\begin{equation}
\rho_{crit}(z)=\rho_{crit,0}E(z)^{1/2}.
\end{equation}
In this paper the normalization factor $f_{T}$ in Eq.(\ref{eq:kt})
is taken to be 1.1. The virial temperature and radius
corresponding to the largest mass of the virialized halos for the
cosmological models are also demonstrated in Table 1.

\begin{table}
\caption{Cosmological models and results}
\begin{center}
\begin{tabular}{||c|c|c|c|c|c||}
\hline\hline
Model &  $\Omega_{m}$ & $\Omega_{\Lambda}$ & M$_{0}({10}^{15}h^{-1}
M_{\odot})$ & T$_{vir}$(kev) & R$_{vir}$($h^{-1}$Mpc)\\
 \hline
 SCDM & 1 & 0 & 20.9 & 65.22 & 4.66\\
 \hline
 LCDM & 0.3 & 0.7 & 9.01 & 30.85 & 4.25\\
 \hline
 OCDM & 0.3 & 0 & 1.62 & 9.83 & 2.40\\
\hline\hline
\end{tabular}
\end{center}
\end{table}
\section{ Conclusion and Discussion}
\hspace{8mm}From Table 1 we can see that the different cosmological
 models
may yield the different result about virial mass, temperature and
radius for
the largest virialized halos. As expected, the result from the flat
 $\Lambda$CDM model
with the largest virialized cluster mass of about $9.01 \times
10^{15}h^{-1}M_{\odot}$
is in good agreement with the mass of $10^{15}h^{-1}M_{\odot}$
for a typical rich cluster,
while that for OCDM model with the mass $1.62 \times 10^{15}
h^{-1}M_{\odot}$ is roughly
consistent with the mass for the typical rich clusters. In addition,
the SCDM model clearly
gives a result in conflict with the other cosmological models and
 results observed. Due to
the accumulative effect of the integration for volume(or redshift)
over the whole space in
the universe, the prediction for virial mass is slightly greater than
the observed one.
 Therefore both SCDM model and OCDM model can be ruled out. On the other
  hand, the measurement
 of luminosity-redshift relation for SN Ia suggests that the expansion
 of the universe is accelerating,
 which indicates the existence of a cosmological constant or dark energy
 possessing negative pressure $p$
 and equation of state $\omega=p/\rho$$^{15,16}$. And the observation
 of curvature of the universe from
 the highest redshift cosmological test-CMB, also reveals that the
 universe is flat, which
  is consistent with the standard inflationary prediction. As a result,
  the precise measurements
  of accelerating expansion of the universe from SN Ia and the spatial
  curvature from the CMB may
   combine to suggest a flat $\Lambda$CDM model with approximately
   $\Omega_{m}=1/3$, $\Omega_{\Lambda}=2/3$
   and $\Omega_{k}=0$$^{17,18}$. In a sense, the obtained largest
   virialized object , which is referred to
    as the complement to the observations of the CMB, SN Ia and the
    large scale structure of the universe,
   may provide a strong support to the present popular $\Lambda$CDM
   model.\\[4mm]
\bigskip
{\bf \Large Acknowledgments}:\\ 
We are grateful for the hospitality of Prof. Xiang-Ping Wu's 
Cosmology group of the National Astronomical
Observatory and Prof. Wu's many useful suggestions.
We thank the anonymous referee for helpful comments.\\
\begin{flushleft}{\bf \Large References}\\
\small
\ 1.\ J.P.Henry and K.A.Arnaud, ApJ. 372,410 (1991).\\
\ 2.\ V.R.Eke, J.F.Navarro and C.S.Frenk, ApJ. 503,569 (1998).\\
\ 3.\  M.Markevitch, W.R. Forman, C.L.Sarazin and A.Vikhlinin,ApJ.
540,27 (1998).\\
\ 4.\ K.M.Ashman, P.Salucci and M.Persic, MNRAS. 260,610 (1993).\\
\ 5.\ W.H.Press, and P.Schechter, Astrophys. J. 187, 425 (1974).\\
\ 6.\ G.Efstathiou et al, MNRAS.235,715 (1988).\\
\ 7.\ R.G.Carlbeng et al, Astrophys. J. 340,47 (1989).\\
\ 8.\ J.A.Peacock et al, MNRAS. 243,133 (1990).\\
\ 9.\ J.R.Bond et al, Astrophys. J. 379,440 (1991).\\
\ 10.\ J.M.Bardeen et al, Astrophys. J. 304,15 (1986).\\
\ 11.\ E.L.Lokas, Y.Hoffman, astroph/0011295.\\
\ 12.\ S.M.Carroll, W.H.Press and E.L.Turner,ARA and A. 30,499 (1992).\\
\ 13.\ G.L.Bryan and M.L.Norman, Astrophys. J. 495,80-99 (1998).\\
\ 14.\ Peebles and P.J.E ,The large-scale structure of the Universe
(Princeton, NJ: Princeton Universe Press) (1980).\\
\ 15.\ A.G.Riess et al, Astrophys. J. 116,1009 (1998).\\
\ 16.\ S.Perlmutter et al, Astrophys. J. 517,565 (1999).\\
\ 17.\ L.Wang et al astro-ph/9901388 (1999).\\
\ 18.\ S.Perlmutter et al, astro-ph/9901052 (1999).\\
\end{flushleft}
\end{document}